\documentclass[12pt,preprint]{aastex}
\usepackage{natbib}

\newcommand{\Hion}{\textrm{H}_2^+}
\newcommand{\Hmol}{\textrm{H}_2}
\newcommand{\Hatom}{\textrm{H}}
\newcommand{\proton}{\textrm{H}^+}
\slugcomment{Accepted for publication in ApJ Suppl.}

\shorttitle{Vibrational state resolved cross sections for $\Hion$}
\shortauthors{Babb}

\begin{document}

\title{State resolved data for radiative association of $\Hatom$ and $\proton$\\
and for photodissociation of $\Hion$}

\author{James F. Babb}
\affil{ITAMP, Harvard-Smithsonian Center for Astrophysics,\\
    MS 14, 60 Garden St., Cambridge, MA 02138-1516}
\email{jbabb@cfa.harvard.edu}

\begin{abstract}
The matrix elements and energies needed to calculate 
vibrational-rotational state resolved cross sections
and rate coefficients for 
radiative association of $\Hatom$ and $\proton$
and for  photodissociation of $\Hion$ 
are presented for applications to simulations of chemistry in 
the early Universe and to stellar atmospheres.
\end{abstract}


\section{Introduction}

Radiative association of $\Hatom$ and $\proton$ produces $\Hion$
in various vibrational-rotational states. 
Recent studies of $\Hmol$ chemistry 
in the early Universe indicated the necessity 
of accounting for the vibrational-rotational state distribution of the $\Hion$ population 
in the absence of 
local thermodynamic equilibrium, which accordingly requires ``state-resolved'', i.e.,
vibrational-rotational state resolved, calculations of 
data 
for the $\Hatom + \proton$ radiative association process and for the inverse process, photodissociation of 
$\Hion$. The absence of such data in the literature was noted
by~\citet{HirPad06}, \cite{CopLonCap11}, and \cite{GloChlFur14}.
In the present paper, the necessary data to calculate
radiative association and photodissociation cross sections 
for each vibrational-rotational level of $\Hion$ are presented.
Cross sections so assembled can then be summed or averaged as required for particular
applications, e.g., rate coefficients for non-local thermal equilibrium models.
In addition to being useful for early Universe chemistry models, 
$\Hion$ photodissociation cross sections enter in models of  the solar atmosphere~\citep{Sta94a,MihIgnSak07},
where $\Hion$ is vibrationally excited~\cite[p.168]{Gla98},
and in models of the atmospheres of DA white dwarfs~\citep{SanKep12}.

The radiative association reaction 
\begin{equation}
\label{ra-rxn}
\Hatom + \proton \rightarrow \Hion (v,N) + h\nu
\end{equation}
and its
inverse process photodissociation
\begin{equation}
\label{pd-rxn}
\Hion (v,N) + h\nu \rightarrow \Hatom + \proton 
\end{equation}
for a photon of energy $h\nu$ 
are related through microscopic reversibility~\citep{MosWu51,MosWu51E,LigRosShu69,HirPad06},
where $(v,N$) denotes the vibrational $v$ and rotational $N$ quantum number
of the molecular ion.

The rate coefficients for~(\ref{ra-rxn}) calculated
by~\citet{Bat51}, \citet{RamPee76}, and \citet{StaBabDal93} are in agreement.
The cross section summed over all bound states $(v,N)$ of $\Hion$ was given by~\cite{StaBabDal93}.

The photodissociation cross sections for (\ref{pd-rxn})  were 
calculated by~\citet{Oks67}, \citet{Dun68a}, \citet{Dun68b}, \citet{Arg74}, \citet{Sta94a}, \citet{LebPre02}, and \citet{MihIgnSak07}.
Where individual data 
for select $(v,N)$ levels and photon energies $h\nu$ is given 
there is general agreement between their calculations.
A tabulation
of the photodissociation cross sections for each of
the vibrational states $(v,1)$ is given in the  report of~\citet{Dun68b}.
Recently, calculations of the photodissociation cross sections were carried out for applications to DA white dwarfs 
by \citet{SanKep12}.

\section{Theory}

The cross section for radiative association~(\ref{ra-rxn}) is~\citep{StaBabDal93}
\begin{equation}
\label{ra-cross-all}
\bar{\sigma}^\mathrm{ra}_N(v,E) = \frac{8}{3}\frac{\pi^2}{\hbar^3c^3} \frac{h^3\nu^3}{k^2} p
 [ N M^2_{v,N-1;k,N} + (N+1)  M^2_{v,N+1;k,N} ] ,
\end{equation}
and $M_{v,N-1;k,N}$ is the matrix element of the transition dipole moment
 between the  continuum wave function of energy $E$ and angular momentum $N$
and the bound vibrational-rotational wave function,
$k^2=2\mu E/\hbar^2$,  $\mu$ is the reduced mass of the colliding system,  
$h\nu$ is the photon energy, 
and $p=1/2$ is the probability of approach in the $2p\sigma_u$ electronic state.

The cross section for photodissociation is~\citep{LebPre02} 
\begin{equation}
\label{pd-cross-all}
\bar{\sigma}^\mathrm{pd}_{v,N} (h\nu) = \frac{8\pi^3\nu}{3c (2N+1)}
 [ (N+1) M^2_{v,N;k,N+1} + N M^2_{v,N;k,N-1} ]  .
\end{equation}

In Eqs.~(\ref{ra-cross-all}) and (\ref{pd-cross-all}) the nuclear spin symmetry
weighting is omitted and it can be included later if the cross sections
are being folded into a temperature distribution~\citep{Sta94a,MihIgnSak07}.
The effect of centrifugal distortion is negligible for the repulsive
$2p\sigma_u$ state and it is a good approximation to replace
$N+1$ and $N-1$ by $N$ in Eqs. (\ref{ra-cross-all}) and (\ref{pd-cross-all}).
Then,
\begin{equation}
\label{ra-cross}
\sigma^\mathrm{ra}_N(v,E) = \frac{8}{3} \frac{\pi^2}{\hbar^3c^3} \frac{h^3\nu^3}{k^2}  p (2N+1) M^2_{v,N;k,N}
\end{equation}
and the cross section for photodissociation is
\begin{equation}
\label{pd-cross}
\sigma^\mathrm{pd}_{v,N} (h\nu) = \frac{4\pi^2 h \nu}{3\hbar c}  M^2_{v,N;k,N} \quad .
\end{equation}
A comparison of~(\ref{ra-cross}) and (\ref{pd-cross}) 
shows that
\begin{equation}
\sigma^\mathrm{ra}_N(v,E) = p \frac{(h\nu)^2}{\mu c^2 E} (2N+1) \sigma^\mathrm{pd}_{v,N} (h\nu) ,
\end{equation}
in accordance with microscopic reversibility~\citep{LigRosShu69}.
Thus, tabulation of $v$, $N$, $E_{v,N}$, $E$, and $M^2_{v,N;k,N}$ 
provides the necessary information for calculations of $\sigma^\mathrm{ra}_N(v,E)$
and $\sigma^\mathrm{pd}_{v,N}(h\nu)$, subject to the 
requirement $\delta (h \nu  - ( |E_{v,N}| + E) )$ 
for either~(\ref{ra-cross}) or (\ref{pd-cross}),
where $E_{v,N}$ is the vibrational-rotational eigenvalue
in the $1s\sigma_g$ state measured with respect to
the dissociation limit (taken to be zero here).

\section{Molecular states}

The Born-Oppenheimer potential energy surfaces for the ground $1s\sigma_g$ state and excited
$2p\sigma_u$ state were calculated using the methods described by~\citet{MadPee71}
and extended asymptotically as in~\citet{StaBabDal93}.
The transition dipole moment was calculated using a variational method~\citep{Bab94}
and extended using the asymptotic formula of~\citet{RamPee73}.
Energy differences and oscillator strengths are in excellent agreement with 
the recent calculations of~\citet{TsoBan10}.

There are 423 bound vibrational-rotational levels for $\Hion$ in the $1s\sigma_g$ state.
In the present work, the Born-Oppenheimer potential is used and 
adiabatic, relativistic, and
radiative corrections are ignored.
It is worth noting that
more precise approaches yield the same number of bound levels~\citep{HunYauPri74,Mos93},
but due to the inclusion of
higher order corrections the resulting eigenvalues differ from those obtained 
using only the Born-Oppenheimer potential.
For the present purposes, the corrections are not needed.

The energy-normalized continuum wave functions on the $2p \sigma_u$ state were
evaluated as in~\citet{StaBabDal93}.

\section{Calculations}

The quantities are calculated at a range of energies
from  the threshold energy $|E_{v,N}|$ for photodissociation  of level $(v,N)$  to
a maximum photon  wavelength of $55~\textrm{nm}$ (22.5~eV or 0.828 a.u.).
Only values of $M^2_{v,N;k,N} > 1\times 10^{-6}$ are listed.

Table~\ref{table} gives
the vibrational quantum number $v$, the rotational quantum number $N$,
the relative energy $E$ in atomic
units of energy (27.2114 eV), 
the eigenvalue $|E_{v,N}|$ in $\textrm{cm}^{-1}$, the photon wavelength $\lambda =c/h\nu$ in nm,
and the value of $M^2_{v,N;k,N}$ in atomic units.

The photodissociation cross section in units of $\textrm{cm}^2$ is~\citep{ElQSta13}
\begin{equation}
\label{pd-tabular}
\sigma^\mathrm{pd}_{v,N} (h\nu) = 2.689 \times 10^{-18} 
      (45.563/\lambda) M^2_{v,N;k,N} \quad \textrm{cm}^2 ,
\end{equation}
with $\lambda$ in nm and $M^2_{v,N;k,N}$ in atomic units.
The radiative association cross section in units of $\textrm{cm}^2$
is~\citep{StaBabDal93}
\begin{equation}
\sigma^\mathrm{ra}_N(v,E) = 1.475\times 10^{-20}   
   (45.563/\lambda)^3 (2N+1) E^{-1} 
   M^2_{v,N;k,N} \quad \textrm{cm}^2,
\end{equation}
with $\lambda$ in nm, $E$ in atomic units, and $M^2_{v,N;k,N}$ in atomic units.

\begin{figure}
%\epsscale{.80}
\plotone{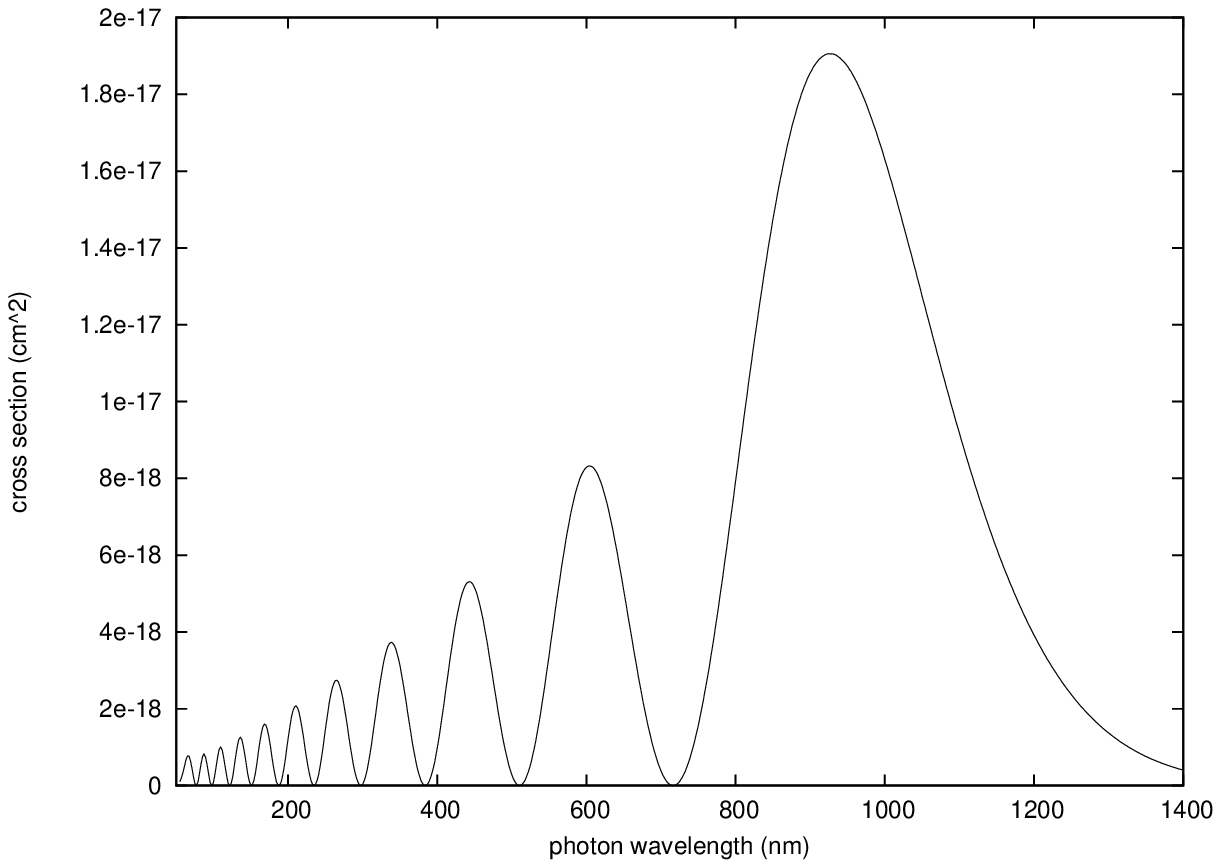}
\caption{For $v=10,N=0$, plot of $\sigma^\mathrm{pd}_{v,N}(h\nu)$ calculated from the data in
Table~(\ref{table}).\label{pd-fig}}
\end{figure}

\citet{Dun68a,Dun68b} calculated the photodissociation cross section for 
$v=0$ to $v=18$ using the potential energy surfaces calculated by
\citet{BatLedSte53}. There are slight differences between the accurate
potential energy surfaces of the present work and the early
calculations of \citet{BatLedSte53}.
Nevertheless, agreement is generally good between the current calculations
and the tabulated values of \citet{Dun68a,Dun68b}.
In Figure~(\ref{pd-fig})
for $v=10,N=0$, a plot of $\sigma^\mathrm{pd}_{v,N}(h\nu)$ calculated from the data in
Table~(\ref{table}) is presented.
The representation is very good and the tabulated data reflect the peaks
arising from the $v=10$ bound state wave function.
The present tabulated data 
can be interpolated using cubic splines, but 
due to the oscillations several intervals should be selected.

\acknowledgments
ITAMP is supported in part by a grant from the NSF to the Smithsonian
Astrophysical Observatory and Harvard University.

%\begin{thebibliography}{}
%\end{thebibliography}

%

%
%%% This table also includes a table comment indicating that the full
%%% version will be available in machine-readable format in the electronic
%%% edition.
%
\clearpage
\begin{deluxetable}{cclrrl}
\tabletypesize{\scriptsize}
%\rotate
\tablecaption{Calculated values of the squared matrix element, $M^2_{v,N;k,N}$, for bound vibrational-rotational levels
($v,N$) of the $1s\sigma_g$ state of $\Hion$ with eigenvalue $|E_{v,N}|$ to a continuum
level of the $2p\sigma_u$ state with energy $E$ for a photon of energy $h\nu$.
In the table, $E$ is in atomic units, $|E_{v,N}|$ is in $\textrm{cm}^{-1}$,
$c/h\nu$ is in nm, and $M^2_{v,N;k,N}$ is in atomic units.
\label{table}}
\tablewidth{0pt}
\tablehead{
\colhead{$v$} & \colhead{$N$} & \colhead{$E$} & \colhead{$|E_{v,N}|$} & \colhead{$c/h\nu$} & \colhead{$M^2_{v,N;k,N}$}
}
\startdata
0 & 0 & 0.7310E+00 & 21375.95 & 55.00 & 0.1333E-03 \\
0 & 0 & 0.7249E+00 & 21375.95 & 55.41 & 0.1687E-03 \\
0 & 0 & 0.7188E+00 & 21375.95 & 55.83 & 0.2126E-03 \\
0 & 0 & 0.7128E+00 & 21375.95 & 56.24 & 0.2667E-03 \\
0 & 0 & 0.7069E+00 & 21375.95 & 56.65 & 0.3329E-03 \\
0 & 0 & 0.7011E+00 & 21375.95 & 57.06 & 0.4138E-03 \\
0 & 0 & 0.6953E+00 & 21375.95 & 57.48 & 0.5122E-03 \\
0 & 0 & 0.6897E+00 & 21375.95 & 57.89 & 0.6312E-03 \\
0 & 0 & 0.6841E+00 & 21375.95 & 58.30 & 0.7747E-03 \\
0 & 0 & 0.6786E+00 & 21375.95 & 58.72 & 0.9469E-03 \\
0 & 0 & 0.6732E+00 & 21375.95 & 59.13 & 0.1153E-02 \\
0 & 0 & 0.6678E+00 & 21375.95 & 59.54 & 0.1398E-02 \\
0 & 0 & 0.6626E+00 & 21375.95 & 59.95 & 0.1689E-02 \\
0 & 0 & 0.6574E+00 & 21375.95 & 60.37 & 0.2033E-02 \\
0 & 0 & 0.6523E+00 & 21375.95 & 60.78 & 0.2438E-02 \\
0 & 0 & 0.6472E+00 & 21375.95 & 61.19 & 0.2914E-02 \\
0 & 0 & 0.6422E+00 & 21375.95 & 61.61 & 0.3470E-02 \\
0 & 0 & 0.6373E+00 & 21375.95 & 62.02 & 0.4118E-02 \\
0 & 0 & 0.6324E+00 & 21375.95 & 62.43 & 0.4872E-02 \\
0 & 0 & 0.6276E+00 & 21375.95 & 62.84 & 0.5743E-02 \\
0 & 0 & 0.6229E+00 & 21375.95 & 63.26 & 0.6750E-02 \\
0 & 0 & 0.6182E+00 & 21375.95 & 63.67 & 0.7907E-02 \\
0 & 0 & 0.6136E+00 & 21375.95 & 64.08 & 0.9235E-02 \\
0 & 0 & 0.6091E+00 & 21375.95 & 64.49 & 0.1075E-01 \\
0 & 0 & 0.6046E+00 & 21375.95 & 64.91 & 0.1248E-01 \\
0 & 0 & 0.6001E+00 & 21375.95 & 65.32 & 0.1445E-01 \\
0 & 0 & 0.5958E+00 & 21375.95 & 65.73 & 0.1668E-01 \\
0 & 0 & 0.5914E+00 & 21375.95 & 66.15 & 0.1920E-01 \\
0 & 0 & 0.5872E+00 & 21375.95 & 66.56 & 0.2204E-01 \\
0 & 0 & 0.5829E+00 & 21375.95 & 66.97 & 0.2523E-01 \\
\enddata
\tablecomments{Table \ref{table} is published in its entirety in the 
electronic edition of the {\it Astrophysical Journal}.  A portion is 
shown here for guidance regarding its form and content.}
\end{deluxetable}
\end{document}